\documentclass[preprintnumbers,amsmath,amssymb,prd, notitlepage,nofootinbib,twocolumn, superscriptaddress]{revtex4}

\usepackage{graphicx}
\usepackage{dcolumn}
\usepackage{bm}
\usepackage{subfigure}
\usepackage{wasysym}

\newcommand{\be}{\begin{equation}}
\newcommand{\ee}{\end{equation}}

\newcommand{\Dtheta}{\Delta \theta}

\newcommand{\Tr}{\mathrm{Tr}}

\newcommand{\biD}{\boldsymbol{D}}
\newcommand{\bitheta}{\boldsymbol{\theta}}
\newcommand{\bithetat}{{\bitheta^{(T)}}} 
\newcommand{\biDtheta}{\boldsymbol{\Dtheta}} 
\newcommand{\ibg}{\boldsymbol{g}} 
\newcommand{\bie}{\boldsymbol{e}} 

\begin{document}

\title{Building Unbiased Estimators from Non-Gaussian Likelihoods \\
  with Application to Shear Estimation} 

\author{Mathew S. Madhavacheril}
\affiliation{Physics and Astronomy Department, Stony Brook University, 
Stony Brook, NY 11794, USA}

\author{Patrick McDonald}
\affiliation{Lawrence Berkeley National Laboratory, One Cyclotron Road,
Berkeley, CA 94720, USA}

\author{Neelima Sehgal}
\affiliation{Physics and Astronomy Department, Stony Brook University, 
Stony Brook, NY 11794, USA}

\author{An\v{z}e Slosar}
\email{anze@bnl.gov}
\affiliation{Brookhaven National Laboratory,
             Blgd 510, Upton NY 11375, USA}

\date{\today}

\begin{abstract}
  We develop a general framework for generating estimators of a given
  quantity which are unbiased to a given order in the difference
  between the true value of the underlying quantity and the fiducial
  position in theory space around which we expand the likelihood. We
  apply this formalism to rederive the optimal quadratic estimator and
  show how the replacement of the second derivative matrix with the
  Fisher matrix is a generic way of creating an unbiased estimator
  (assuming choice of the fiducial model is independent of data). Next we
  apply the approach to estimation of shear lensing, closely following
  the work of Bernstein and Armstrong (2014). Our first order
  estimator reduces to their estimator in the limit of zero shear, but
  it also naturally allows for the case of non-constant shear and the
  easy calculation of correlation functions or power spectra using
  standard methods.  Both our first-order estimator and Bernstein and
  Armstrong's estimator exhibit a bias which is quadratic in true
  shear. Our third-order estimator is, at least in the realm of the
  toy problem of Bernstein and Armstrong, unbiased to 0.1\% in
  relative shear errors $\Delta {\ibg}/{\ibg}$ for shears up to
  $|{\ibg}|=0.2$.
\end{abstract}

\maketitle

\section{Introduction}
Unbiased estimators are recipes for producing an estimate of a
quantity which, averaged over many realizations of the data from the
same underlying model, will average towards the true value of the
quantity we seek to measure (assuming the averaging is unweighted, 
or symmetrically weighted). 

A typical example of where unbiased estimators might be useful is the
estimation of cosmic shear. One can write the complete likelihood for
the observed galaxy image given the parameters of the galaxy
model. Such a model might include parameters describing the intrinsic
ellipticity of the galaxy, its size, etc. and also the quantities that
one wants to measure, such as shear. In general, the resulting
likelihood will be very non-Gaussian, i.e. it cannot be usefully
described by the position of maximum likelihood and the second
derivative matrix around that point in parameter space.  In order to
carry out an analysis in an unbiased manner, one would need to
propagate the full likelihood shape in the subsequent analysis of the
data. This is prohibitive in the limit of millions of galaxies whose
shear one hopes to measure in forthcoming surveys. One could attempt
to maximize the likelihood for each individual galaxy, but this
typically leads to wrong answers -- since galaxies are round on
average, a given galaxy might be best explained as a result of massive
shearing of an intrinsically round galaxy. But we know that a model
with a shear of say $0.3$ does not make much sense for a typical field
galaxy. In \cite{2014MNRAS.438.1880B} (BA14 hereafter), the authors
have argued for the expansion of the marginalized likelihood around
zero shear, i.e. compressing the likelihood to the value of the first
and second derivatives of the log-likelihood expanded around zero
shear. The fact that the likelihood for each individual galaxy is
highly non-Gaussian does not matter.  Since the shear is small, when
many log-likelihoods are added (i.e. likelihoods combined), the
resulting likelihood has to collapse to a Gaussian by the central
limit theorem. For such a collapsed likelihood, one can use a
Newton-Raphson step (using the first and second derivatives of the
combined likelihood) to calculate an estimate of the underlying
shear. In BA14, the authors show that this method works on a toy
example (also employed later in this paper), and
\cite{2014arXiv1403.7669S} demonstrates that it also performs as
expected in more realistic settings (e.g. working with real pixelated
galaxy images, but still using simulations).

However, one caveat to the method discussed above is that, in its
simplest incarnation presented in BA14, it only works when the shears
of all galaxies are assumed to be the same - something that is clearly
not true in reality. The method requires the likelihood to be combined
for a sufficiently large number of galaxies so that central limit
theorem ensures we can get a sufficiently Gaussian shear estimate for
the ensemble.  Therefore, in order to calculate a correlation function
or a power spectrum, one can either perform shear averaging in 
cells where the shear can be roughly assumed constant,
or, alternatively, attempt to appropriately weight the estimates using cells in Fourier space to
recover individual Fourier modes of the shear field (see Section 2.2
in \cite{2014MNRAS.438.1880B}).

In this paper, we develop a related scheme.   In contrast to the BA14 method, 
where one does not recover an estimate of the shear of a single galaxy, 
the method in this paper does return an
unbiased estimate of the shear for each galaxy. For each individual
galaxy, we make no guarantee as to the probabilistic distribution for
the error $\boldsymbol{\epsilon}=\tilde{{\ibg}}-{\ibg}$ (where $\tilde{\ibg}$
is the shear estimate and ${\ibg}$ is the true shear), except that
$\left<\boldsymbol{\epsilon}\right>=0$, where the average is over all possible
realizations of the data.  Again, while the error properties for
a single galaxy are unknown, they must converge to a normal
distribution when many galaxies are considered by the central limit
theorem.  An important advantage in returning the shear of each galaxy, is that we are now not limited to the 
case of constant shear and can calculate any correlation function using these
estimates, since it is trivial to show, for example, that $\left<
  \tilde{\ibg}_1 \tilde{\ibg}_2 \right>= {\ibg}_1 {\ibg}_2$,
where indices 1 and 2 correspond to two galaxies, $\tilde{\ibg}$ corresponds to the estimated shear, and
${\ibg}$ corresponds to the true shear.

In section \ref{sec:formalism}, we develop the formalism used in this
work, which is completely general and independent of any particular
inference problem. It will turn out that in general, an estimator can
be constructed that is unbiased to a certain order in the difference
between the true and assumed fiducial values for the theory
parameters. In Section \ref{sec:optim-quadr-estim}, we re-derive the
optimal quadratic estimator in our formalism, and in Section
\ref{sec:shear-estimation}, we apply our formalism to the toy problem
of BA14.

\section{Formalism}
\label{sec:formalism}

\newcommand{\lL}{\mathcal{L}}
Consider a general likelihood function $L(\biD;\bitheta)$, which
is a function of a vector of $N$ theory parameters $\bitheta$ and a
vector of $M$ observable data values $\biD$.\footnote{We follow
  standard notation where vectors and matrices which are not
  explicitly indexed are denoted with bold-face italic
  font and bold-face roman fonts respectivelly.} We will denote the log
likelihood as $\lL=\log L$. The likelihood is normalized as
\begin{equation}
  \int L d^M\biD = \int e^\lL d^M \biD=1.
\label{eq:3}
\end{equation}
The above is true for \emph{any} set of theory parameters $\bitheta$.  We
will write the average of any quantity over the likelihood at theory
parameter $\bitheta$ as
\begin{equation}
  \left<X(\biD;\bitheta') \right>_\theta = \int X(\biD;\bitheta') e^{\lL(\biD;\bitheta)} d^M\biD 
\label{eq:3x}
\end{equation}
Note that the function $X$ can in general be a function of both data and
the theory parameters, but the resultant average  $\left<X(\biD;\bitheta')
\right>_{\bitheta}$ is a function of $\bitheta$ and $\bitheta'$, but not $\biD$.
Let us denote the derivative with respect to the theory parameters
with a comma, i.e.  $\lL_{,i}=\frac{\partial \lL}{\partial \theta_i}$.
The first derivative $\lL_{,i}$ is a vector of size $N$, the second
derivative $\lL_{,ij}$ is a symmetric matrix of size $N\times N$, etc.

 Taking $n$ derivatives of Equation \eqref{eq:3} with respect to
 theory parameters, we find that
\begin{equation}
  \left<^{n}\mathbf{U}(\bitheta) \right>_{\bitheta} = 0\\
\label{eq:4}
\end{equation}
where we have introduced the shorthand notation
\begin{eqnarray}
  ^{1}U_i &=& \frac{L_{,i}}{L} = \lL_{,i} \label{eq:u1}\\
  ^{2}U_{ij} &=& \frac{L_{,ij}}{L} = \lL_{,ij}+\lL_{,i}\lL_{,j}\\
  ^{3}U_{ijk} &=& \frac{L_{,ijk}}{L} = \lL_{,ijk} + \lL_{,ij}\lL_{,k}  + {\rm cyc} + \lL_{,i} \lL_{,j} \lL_{,k}\\
  ^{n}\mathbf{U} &=& \frac{1}{L} \frac{\partial^n L}{\partial \bitheta^n} =
  \frac{\partial}{\partial \bitheta} {^{n-1}\mathbf{U}} + {^{n-1}\mathbf{U}}{^1\mathbf{U}}
\end{eqnarray}
Note that Equation \ref{eq:4} only holds when both the $\bitheta$ inside the brackets and 
outside the brackets are the same.
In general, however, in Equation \ref{eq:3x}, the $\bitheta '$ appearing in $X$ need not be at the same 
position in theory space as the $\bitheta$ appearing in $L(\biD;\bitheta)$. 

The first of the above equations, namely $\left<\lL_{,i}\right>=0$ has a
very clear physical interpretation. It is telling us, that if one chooses
a theoretical model specified by $\bithetat$, generates a set of observed
data points $\biD$ given that model, calculates the first
derivative of the log-likelihood at the true model value
$\lL_{,i}(\biD;\bithetat)$, and then averages this quantity over all possible
realizations of the data, then the result will be zero.  In fact, this must intuitively
be so: if one has access to many realizations of the data from the
same theory available, multiplying likelihoods (or equivalently adding
log-likelihoods) will result in a Gaussian likelihood that will become
increasingly tightly centered on the true value. In the limit of the
infinite number of data realizations, it becomes a delta function at
the true value.

Of course, this is not very helpful, since if we knew the true value,
we would not need to measure it. So, let us assume that the true value 
is at some nearby position $\bithetat=\bitheta + \Delta \bitheta$. If we expand
the likelihood around $\bitheta$ (note that we are not expanding around
the true model, but around a chosen fiducial model), we find
\begin{equation}
e^{\lL (\bithetat)} =  e^{\lL(\bitheta)}\left(1 + 
  {\sum_{n=1}^\infty} \frac{1}{n!}  {^n \mathbf{U}}(\bitheta) \Delta \bitheta^n \right).
\label{eq:8}
\end{equation}
Note that the $n$-th term in the Taylor expansion is a product of $^nU$,
which has $n$ indices, with $\Delta \bitheta^n=\Delta \theta_i \Delta
\theta_j \ldots \Delta \theta_l$, which also has $n$ indices.

Substituting the right side of Equation \ref{eq:8} into Equation
\ref{eq:3x} gives
\begin{equation}
\left<^m\mathbf{U}(\bitheta)\right>_{\bithetat} =  \sum_{n=1}^{\infty}
\frac{1}{n!}  {^{mn}\mathbf{W}} \Delta \bitheta^n,
\label{eq:rax}
\end{equation}
where 
\begin{equation}
  ^{mn}\mathbf{W} = \left<^m\mathbf{U}  ^n\mathbf{U} \right>_{\bitheta}
\end{equation}
Note that the $^{mn}\mathbf{W}$ object has $m+n$ indices and is only a function
of $\bitheta$, not $\biD$.  We see that quantities $^n\mathbf{U}$ are special. They
average to zero, if we are sitting on a true model
($\left<^{n}\mathbf{U}(\bithetat) \right>_{\bithetat} = 0$ as in Equation \ref{eq:4}
since $\Delta \bitheta = 0$ when $\bitheta = \bitheta^T$).  However, as the
true model slips away, those averages analytically respond to the
difference between the true and the fiducial model (as described by
Equation \ref{eq:rax}).

The motivation for all this may be opaque at this point. The important thing
to recognize is that both $^m \mathbf{U}(D;\bitheta)$ and $^{mn}\mathbf{W}(\bitheta)$ are things that
we can compute, given data and a {\rm choice} of fiducial parameters 
$\bitheta$, so estimators of $\bitheta^T$, or equivalently 
$\Delta \bitheta = \bitheta^T-\bitheta$,  can be constructed out of them. 

\subsection{First-order estimator}

Before proceeding, we note that 
\begin{equation}
  ^{11}W_{ij} = \left<\lL_{,i} \lL_{,j} \right> = -\left<\lL_{,ij} \right> = F_{ij}
\end{equation}
is the Fisher matrix (where we have used Equation \ref{eq:4} for $n=2$).

Our first-order estimator comes from inspecting Equation \ref{eq:rax}
for the case when  $\Delta \bitheta$ is sufficiently small that the series can be
truncated at the first order. We can write down the ansatz
\begin{equation}
  \boldsymbol{E}_1 = (^{11}\mathbf{W})^{-1}\, {^1\mathbf{U}} = F^{-1}_{ij} \lL_{,j}.
\label{eq:e1}
\end{equation}
Plugging this solution back into Equation \ref{eq:rax} and remembering that
$^{mn}\mathbf{W}$ is not a function of $\biD$ gives
\begin{eqnarray}
  \left< \boldsymbol{E}_1 \right>_{\bithetat} &=& (^{11}\mathbf{W})^{-1} \left<^1\mathbf{U} \right>_{\bithetat} \\
  &=& \Delta \bitheta_1  + \frac{1}{2} \left(\mathbf{F}^{-1}\right)\, ^{12}\mathbf{W} \Delta \bitheta ^2 + \ldots
\label{eq:e1_thetat}
\end{eqnarray}
This estimator is thus unbiased to quadratic order in
$\Delta \bitheta$. Note that since $\bitheta$ is known (i.e. it is the
assumed fiducial model), we can simply add it to $\boldsymbol{E}_1$ to convert an
estimator of $\Delta \bitheta$ to an estimator of $\bithetat$.
The variance of the estimator is given by
\begin{equation}
  {\rm Var}(\boldsymbol{E}_1) = \mathbf{F}^{-1} + \mathbf{F}^{-1} \mathbf{F}^{-1} \Delta \bitheta \left<^1\mathbf{U} ^1\mathbf{U} ^1\mathbf{U} \right> +\ldots,
\end{equation}
where the  contraction of indices goes as $\left[\mathbf{F}^{-1} \mathbf{F}^{-1} \Delta \bitheta
  \left<^1\mathbf{U} ^1\mathbf{U} ^1\mathbf{U} \right>\right]_{ij} = F^{-1}_{ik} F^{-1}_{jl}
\Delta \bitheta_m \left<^1U_k ^1U_l ^1U_m \right>$.  Thus, given the
Cramer-Rao bound, we have shown that this estimator is unbiased to
quadratic order in $\biDtheta$ and optimal to first order in $\biDtheta$.

\subsection{Higher-order estimators}

To construct higher-order estimators, we need to use  higher order
$\mathbf{U}$s. A quantity of the form 
\begin{equation}
  \boldsymbol{E}_o = \sum_{m=1}^o ({^m\mathbf{A}}) ({^m\mathbf{U}}),
\end{equation}
where $^m\mathbf{A}$ is a $m+1$ index object (indices of the parameter derivatives,
i.e., see Eq. \ref{eq:u1}, etc.), 
will have the mean given by
\begin{equation}
  \left<\boldsymbol{E}_o\right>_{\bithetat} =  \sum_{n=1}^{\infty} \frac{1}{n!}
  \left( \sum_{m=1}^o ({^m\mathbf{A}}) ({^{mn}\mathbf{W}})  \right) \biDtheta^n 
\end{equation}

For a given order $o$, the weights $\mathbf{A}$ can be arranged so that the
pre-factor to $\Delta\bitheta$ is unity and the prefactor to $\delta\bitheta^2$ and
higher are zero up to order $o$. For a concrete example see Section
\ref{sec:shear-estimation} and Appendix \ref{sec:general-3rm-rd}. One
should note that higher order estimators, in general, have higher
variance with respect to the first-order estimator, however, they are
less biased.

Finally, we note that while this construction uniquely specifies one
possible estimator unbiased to a given order, it is clearly not
unique, since one could imagine constructing estimators that are
non-linear in $\mathbf{U}$ quantities and which might, in general, perform
better or worse than this one. We leave investigation of these
questions to future work.

\subsection{A note on iterations}

Since the first-order estimator is accurate to $\Delta\bitheta$, one might be
tempted to simply iterate: start with a first-order estimator, move by
$\Delta\bitheta$, do another iteration there, etc. Note, that such a process
will in general take you to the maximum likelihood point, since the
first-order estimator resembles a Newton-Raphson step. 

It is known that maximum likelihood is not, in general, an unbiased
estimator (although it often happens to be, e.g. for mean and variance
of a Gaussian likelihood). We provide a concrete example in Appendix
\ref{sec:biasml}.  So, why does an iterative process not produce an
unbiased estimate?  The subtlety lies in the fact that the above
derivation assumes that the fiducial $\bitheta$ was chosen without
knowing about the data.  Any iterative process  necessarily breaks
this assumption. Thus, to estimate the mean of an estimator after
several iterations, one would need to average not only over possible
realizations of the data, but also over all possible ``paths'' in the
theory space that a certain iterative process might take.  So, in
general, one should use a higher-order estimator to improve on the
accuracy of the first-order estimator, instead of iterating.

Of course, we expect that the bias due to iteration will be small when
the signal-to-noise is high, so that this will not matter in practice
in those cases.

\section{Optimal quadratic estimator}
\label{sec:optim-quadr-estim}

For completeness, we begin by applying the above formalism to a common
inference problem.  To construct an optimal quadratic estimator
\cite{1998PhRvD..57.2117B,1998ApJ...503..492S,2000ApJ...533...19B},
we start with the data vector $\biD_i$, with zero mean
($\left<\biD\right>=0$), whose covariance can be modeled as
\begin{equation}
  \mathbf{C} = \left<\biD\biD^T\right> =  \mathbf{N}+\theta_i \mathbf{S}_i.
\label{eq:coqe}
\end{equation}
Here $\theta_i$ are some parameters describing the two-point function
of the data, i.e. power spectrum or correlation function bins,
$\mathbf{S}_i$ is the response of the covariance to a change in the value of
$\theta_i$, and $\mathbf{N}$ is assumed to be a known ``noise'' matrix.

Ignoring constant terms, the log-likelihood can be written as
\begin{equation}
\lL = -\frac{1}{2} \log \det \mathbf{C} - \frac{1}{2} \biD^T \mathbf{C}^{-1}\biD.
\end{equation}
In our notation, we have
\begin{equation}
  ^1U_i = -\frac{1}{2} {\rm Tr}\left(\mathbf{C}^{-1}\mathbf{S}_i\right) + \frac{1}{2} {\rm Tr} \left(\biD^T\mathbf{C}^{-1}\mathbf{S}_i\mathbf{C}^{-1}\biD\right).
\end{equation}
A brief calculation gives
\begin{equation}
  \left<^1U_i\right>_{\bithetat} = \frac{1}{2} {\rm Tr} \left(\mathbf{C}^{-1}\mathbf{S}_i
    \mathbf{C}^{-1}\mathbf{S}_j\right) \Dtheta_j
\end{equation}
where we have used $\mathbf{C}(\bitheta^T) =  \left<\biD\biD^T\right>_{\bithetat} =  \mathbf{N}+\bitheta^T_i \mathbf{S}_i = \mathbf{C}(\bitheta) + \Dtheta_i \mathbf{S}_i$,
and hence
\begin{eqnarray}
\left<{\rm Tr} \left(\biD^T\mathbf{C}(\bitheta)^{-1}\mathbf{S}_i\mathbf{C}(\bitheta)^{-1}\biD\right)\right>_{\bithetat} = \\
{\rm Tr} \left(\mathbf{C}(\bitheta)^{-1}\mathbf{S}_i\mathbf{C}(\bitheta)^{-1}(\mathbf{C}(\bitheta) + \Dtheta_j \mathbf{S}_j)\right).
\end{eqnarray}
It follows that
\begin{eqnarray}
  ^{11}\mathbf{W} &=& F_{ij} = \frac{1}{2}\Tr \left(\mathbf{C}^{-1} \mathbf{S}_{i} \mathbf{C}^{-1} \mathbf{S}_{j} \right)\\
  ^{n1}\mathbf{W} &=& 0 \,\  \mathrm{for}\,\   n>1. 
\end{eqnarray}
Plugging these into Equation \eqref{eq:e1}, we recover the standard optimal
quadratic estimator
\begin{equation}
  \boldsymbol{E}_1 = \frac{1}{2}\left[\mathbf{F}^{-1}\right]_{ij} \left[ \biD^T\mathbf{C}^{-1}\mathbf{S}_j\mathbf{C}^{-1}\biD - b_j \right],
\label{eq:estx}
\end{equation}
where $b_i={\rm Tr}\left(\mathbf{C}^{-1}\mathbf{S}_i\right)$.
We have therefore recovered the standard optimal quadratic
estimator and at the same time shown that it is unbiased at all
orders.  The fact that $^{n1}\mathbf{W} = 0$ for $n>1$ implies that this
estimator is unbiased at all orders. Additionally, it can be shown
that this estimator is unbiased regardless of the assumption of a
Gaussian likelihood by calculating the expectation value of the above
equation.  However, this is not directly connected to the framework
here.  (Again, we note that the expectation value proving that the
standard quadratic estimator is unbiased assumes that the covariance
matrix that appears in it does not depend on the data, 
but this assumption is invalidated by iteration.)

These beautiful properties are, of course, crucially dependent on the
theory covariance matrix being linear in theory parameters in Equation
\eqref{eq:coqe}. Fortunately, this is the case in the standard for
measurement of the power spectrum and its linear cousins such as
correlation function. If this is not the case, one can always Taylor
expand around fiducial model and the derivation is then the same with
$\mathbf{N}$ replaced with $\mathbf{N}+\mathbf{C}_{\rm fid.}$, but the
estimator is then only valid within the accuracy of this
approximation.

While this result is not new, it is important to put this into
context. Traditionally, quadratic estimators are often cast as a
Newton-Raphson step towards higher likelihood (see
e.g. \cite{Dodelson-Cosmology-2003}), but here one must remember that,
if the goal is simply function maximization, the true second
derivative may not give the best performance. Numerical work has shown
that performing a Newton-Raphson step with the true second derivative
instead of the Fisher matrix can be an order of magnitude slower in
convergence to the maximum (e.g., when starting power spectrum
parameters are far below the true value).  This is because the true
second derivative and the Fisher matrix are increasingly different as
we move away from the true position in parameter space.  Since the
Fisher matrix estimate is unbiased, one might expect that anything
that deviates from the Fisher estimate must be suboptimal with slower
convergence (strictly speaking, being unbiased does not guarantee
faster convergence if the scatter around the mean is larger but in
practice we do not expect this to happen).  We note however, that even
though an estimate is unbiased when starting with a model that is a
very poor match to the true model, the uncertainties based on a Fisher
matrix will nevertheless be grossly misestimated.

\begin{figure*}
  \centering
  \includegraphics[width=0.24\linewidth]{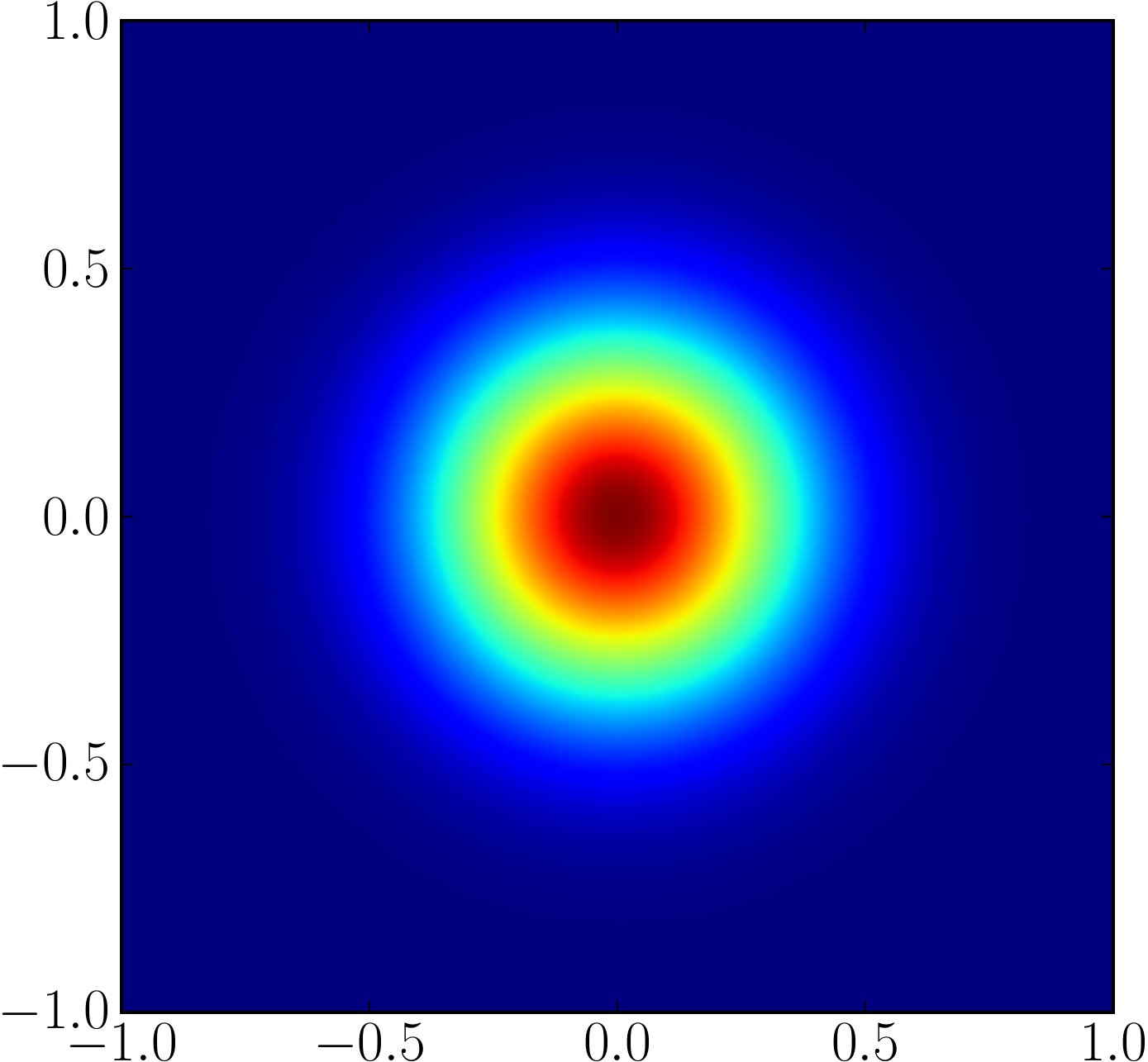}
  \includegraphics[width=0.24\linewidth]{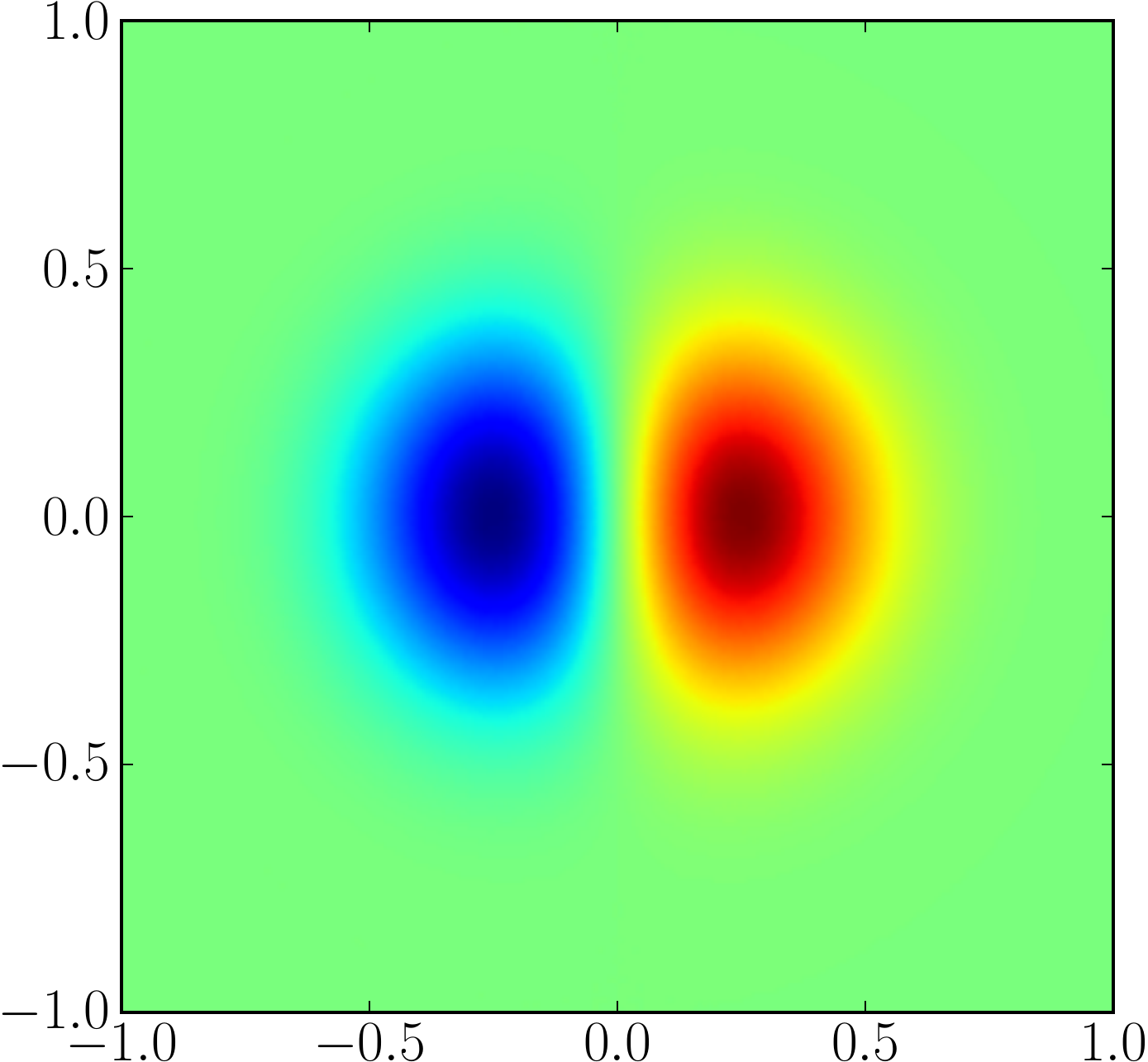}
  \includegraphics[width=0.24\linewidth]{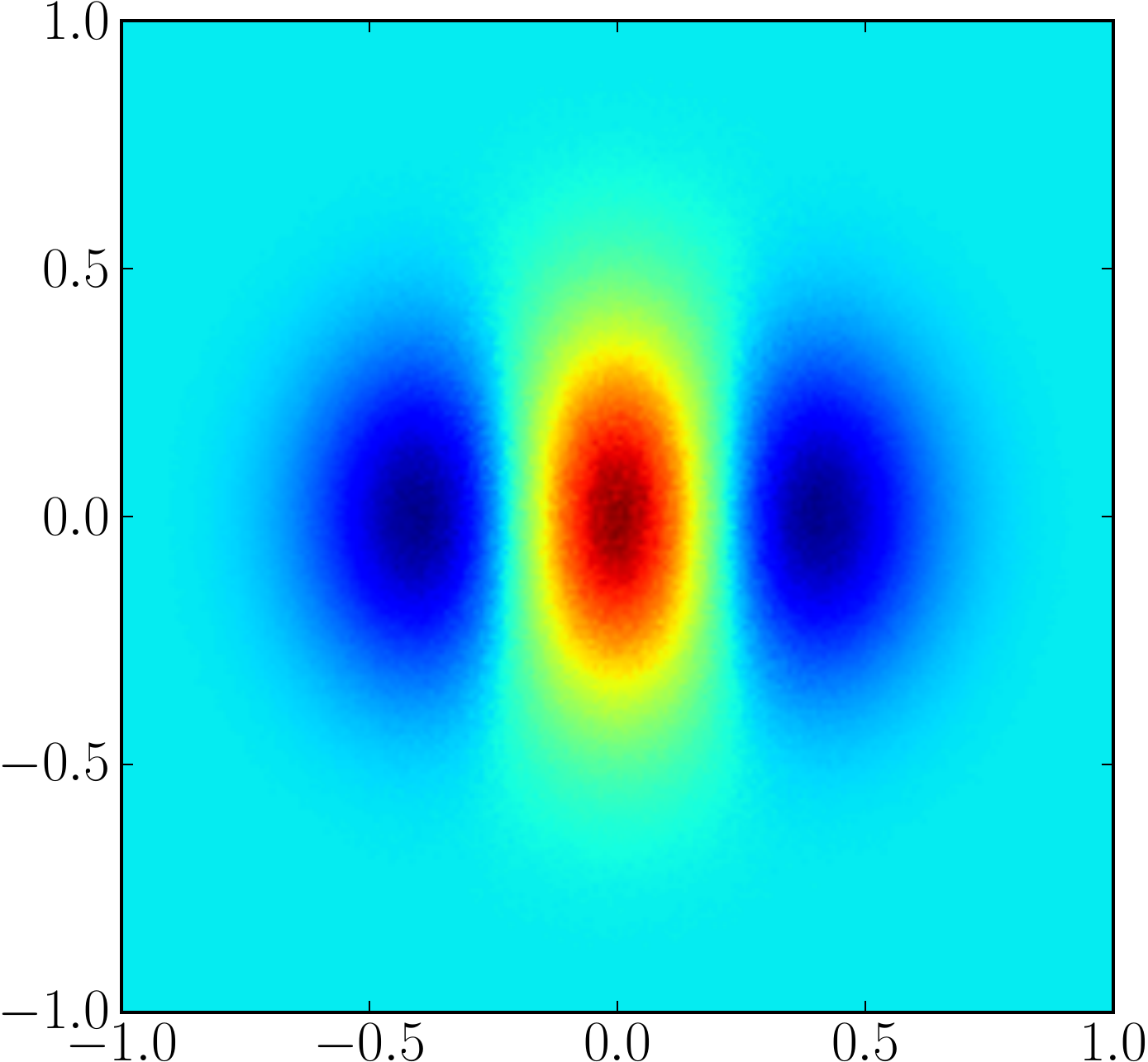}
  \includegraphics[width=0.24\linewidth]{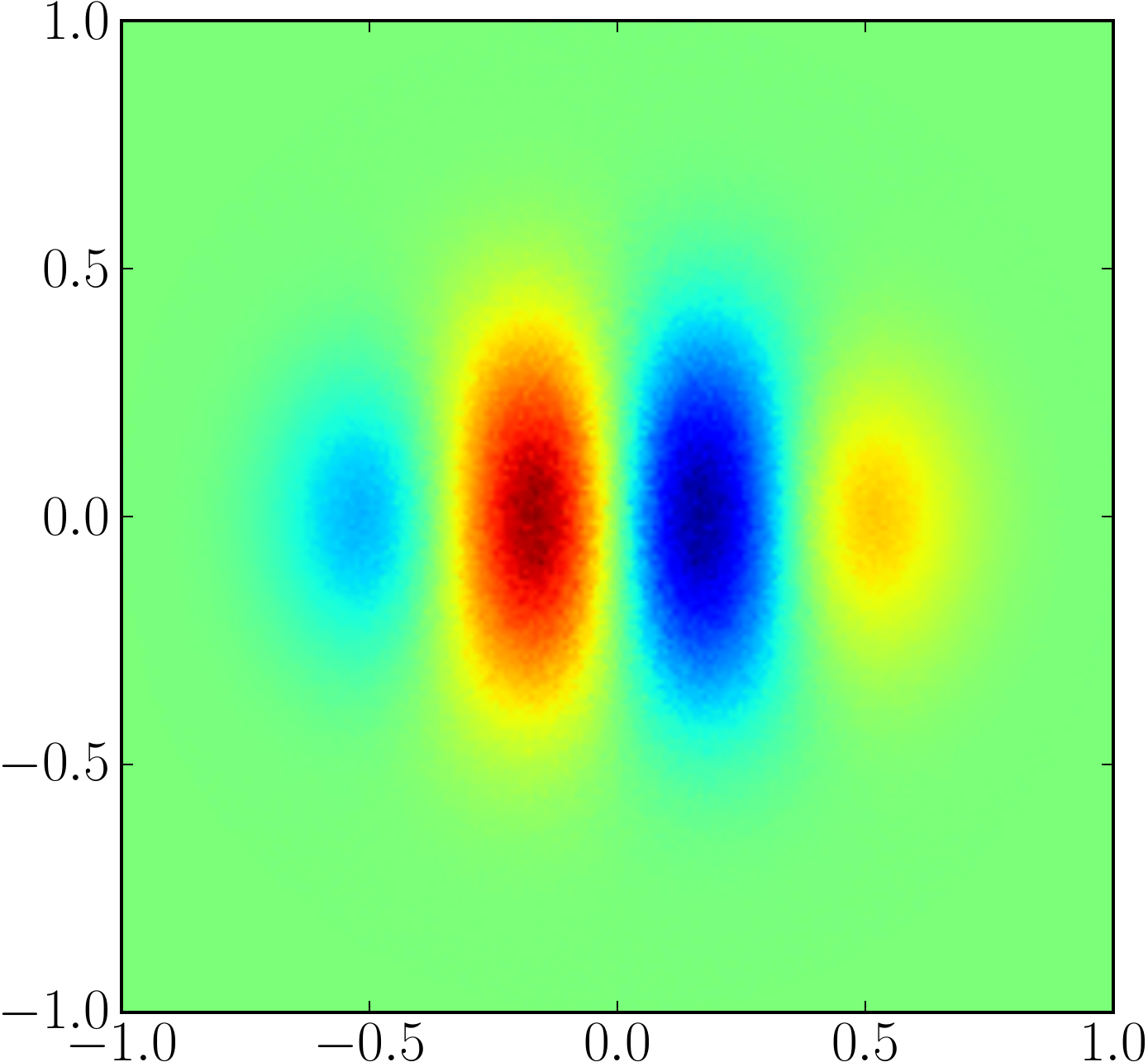}

  \caption{The $i$-th derivative of the likelihood with respect to
    $g_1$ for the posterior distribution at zero shear, where i=0,1,2,3 for
    the toy model described in the text. The x and y axis are the
    measured ellipticities for $e_1$ and $e_2$ respectively, and the color bar saturates
    positively at red and negatively at blue.}
  \label{fig:1}
\end{figure*}

\begin{figure}
\includegraphics[width=\linewidth]{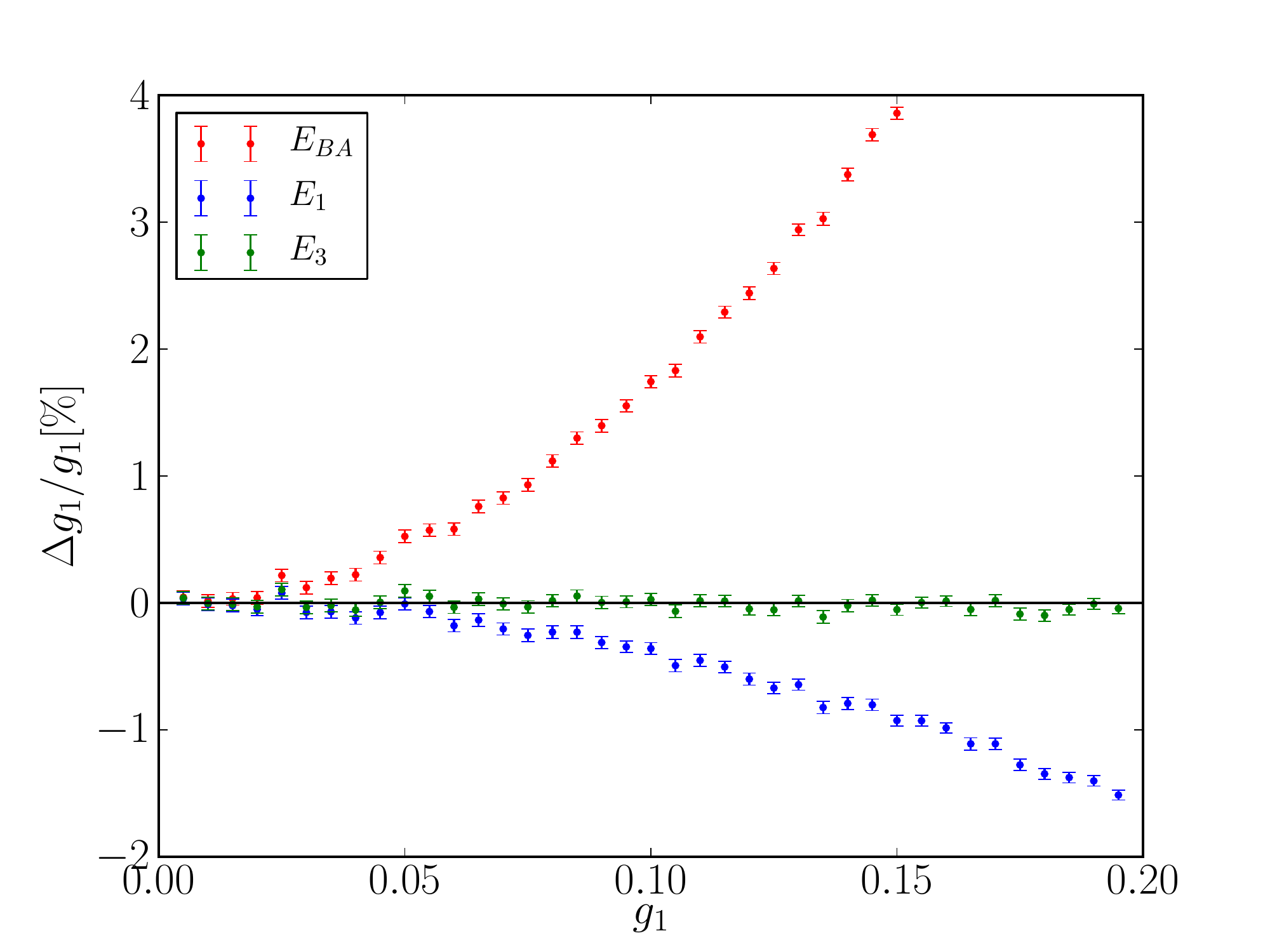}
\caption{The relative biases in the recovered $g_1$ as a function of
  the input $g_1$, with input $g_2$ held at zero. For the $E_1$ and
  $E_3$ estimators, the error was calculated from the variance in estimates, while
  for the $E_{AB}$ estimator, it was assumed to be given by the inverse of the second derivative
  of the posterior.}
\label{fig:2}
\end{figure}

\begin{figure}
\includegraphics[width=\linewidth]{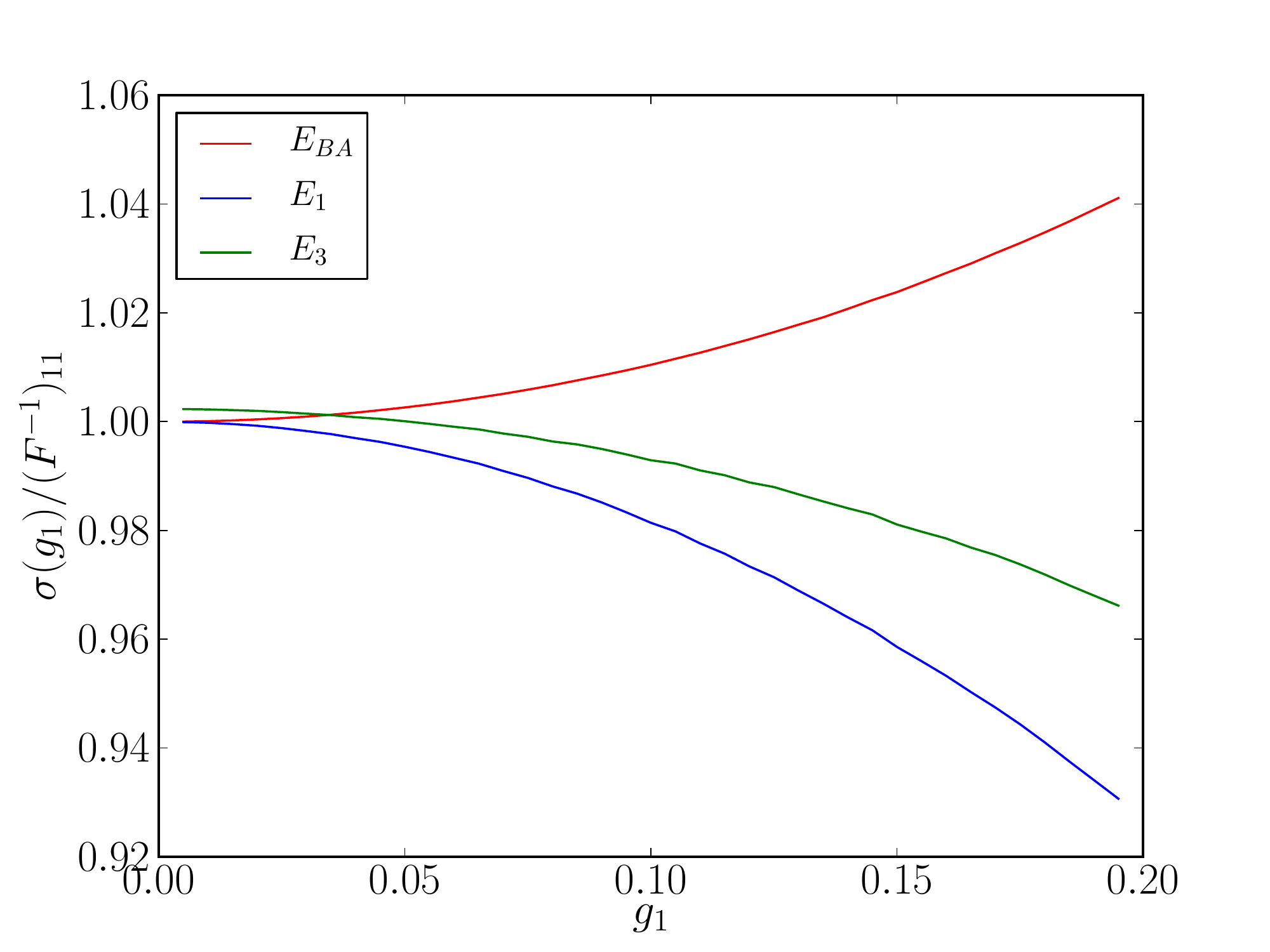}
\caption{The error of estimators relative to the Fisher matrix
  prediction at zero shear. For the $E_1$ and $E_3$ estimators, the error was
  calculated from the variance in estimates, while for the $E_{AB}$ estimator, it
  was assumed to be given by the inverse of the second derivative of the posterior.}
\label{fig:3}
\end{figure}

\section{Shear estimation}
\label{sec:shear-estimation}

To apply the formalism above to the problem of shear estimation, we take
as a starting point work in \cite{2014MNRAS.438.1880B}.  We describe the
likelihood for shear, $L(\ibg)$, through its derivatives
at zero shear as:
\begin{eqnarray}
P &=& L(\biD|\ibg=0) \\
\mathbf{Q} &=& \nabla_{\ibg} L(\biD|\ibg) |_{\ibg=0} \\
\mathbf{R} &=& \nabla_{\ibg} \nabla_{\ibg} L(\biD|\ibg) |_{\ibg=0} \\
\mathbf{S} &=& \nabla_{\ibg} \nabla_{\ibg} \nabla_{\ibg} L(\biD|\ibg) |_{\ibg=0} 
\end{eqnarray}
BA14 expand to second order, but we generalize to third. Note that
theory parameters here are the two components of shear, and we will use
$\ibg$ and $\bitheta$ interchangeably below.  Derivatives of log
likelihood (at zero shear) are thus given by
\begin{eqnarray}
  \lL_{,i} &=& \frac{Q_i}{P} \\
  \lL_{,ij} &=& \frac{R_{ij}}{P} - \frac{Q_iQ_j}{P^2} \\
  \lL_{,ijk} &=& \frac{S_{ijk}}{P} -  \left( \frac{R_{ij}Q_k}{P^2} + {\rm
        cyc}\right) + 2\frac{Q_i Q_j Q_k}{P^3},
\end{eqnarray}
and the $U$ quantities are given simply by
\begin{eqnarray}
  ^1U_i  &=& \frac{Q_i}{P} \\
  ^2U_{ij}  &=& \frac{R_{ij}}{P} \\
  ^3U_{ijk}  &=& \frac{S_{ijk}}{P}.
\end{eqnarray}

BA14 advocate calculating the above quantities for
each galaxy.  If all galaxies have the same shear, the total  probability
can be calculated by summing derivatives of the log likelihood. For a
sufficient number of galaxies, the likelihood collapses to a Gaussian
and the shear can be estimated as 
\begin{equation}
  \boldsymbol{E}_{BA} = -\left( \sum \lL_{,ij}\right)^{-1} \left( \sum \lL_{,j} \right)
\end{equation}
For a sufficiently large number of galaxies $N_g$, the sum of second
derivatives will approach
\begin{eqnarray}
  \sum_1^{N_g} \lL_{,i} &\rightarrow& N_g \left<
    \lL_{,i}\right>_{\bithetat} \\
  \sum_1^{N_g} \lL_{,ij} &\rightarrow& N_g \left<
    \lL_{,ij}\right>_{\bithetat}
\end{eqnarray}
Summing the first and second derivatives of the log likelihood is akin to
averaging over the true distribution.
Therefore, in the limit of an infinite number of galaxies, the
estimator will give
\begin{equation}
  \left< \boldsymbol{E}_{BA} \right>_{\bithetat} = -\left (\left< \lL_{,ij}(\bitheta)\right>_{\bithetat}\right)^{-1} \left< \lL_{,j} \right>_\bithetat
\end{equation}
Note that this is subtly different from our estimator, which uses the
Fisher matrix, $F_{ij} = -\left< \lL_{,ij}(\bitheta)\right>_{\bitheta}$, which is the mean of the second derivative of the log likelihood \emph{assuming zero shear}:
\begin{equation}
  \left< \boldsymbol{E}_1 \right>_{\bithetat} = -\left (\left< \lL_{,ij}(\bitheta)\right>_{\bitheta}\right)^{-1} \left< \lL_{,j} \right>_\bithetat
\end{equation}

\subsection{Toy model}
To test the above ideas, we use the same toy model that was used
in BA14. 
\newcommand{\ve}{\boldsymbol{e}}
We draw a source ellipticity from an isotropic unlensed
distribution with probability distribution given by
\begin{equation}
  P(|\bie^i|) \propto (1-|\bie^i|^2)^2 \exp\left(-\frac{|\bie^i|^2}{2\sigma^2_p} \right)
\label{eq:toy1}
\end{equation}
for the magnitude of the ellipticity and a random orientation.  
The effect of shear is most easily expressed
if we cast the intrinsic  ellipticity and shear as complex vectors
$\bie^i=e_1^i+ie_2^i$ and $\ibg = g_1+ig_2$. Then the sheared
ellipticity vector is given by 
\begin{equation}
  \bie^s = \frac{\bie^i-\ibg}{1-\ibg^*\bie^i}.
\label{eq:toy2}
\end{equation}
Finally, we add random Gaussian noise to obtain the observed
ellipticity $\bie^o$:
\begin{equation}
  \bie^o = \bie^s + \boldsymbol{\epsilon},
\label{eq:toy3}
\end{equation}
where each component of $\boldsymbol{\epsilon}$ is drawn from a truncated Gaussian
with variance $\sigma_n$ ensuring that $|\ve^o|<1$ (in practice random
realizations of noise are added to $\ve^s$ until $|\ve^o|<1$ is satisfied).
In this work we limit ourselves to the example of $\sigma_p=0.3$ and
$\sigma_n=0.05$.

\subsection{Third-order estimator}

It is clear that at least in the case of this particular problem,
symmetry ensures that the second order correction to the estimator
vanishes if one expands around zero shear. There are several ways to
see this. First, given that shear is a spin-2 quantity, the lowest
order scalar one can make is $|\ibg|^2$ and therefore, one expects the
lowest-order correction to an estimate of $\ibg$ to scale as $\ibg
|\ibg^2|$, which is third order in $\ibg$. Second, if one only estimates
$g_1$, it is natural to expect that the correction to $g_1$ must be
the same and of opposite sign to the correction to $-g_1$ --
estimation of shear must be symmetric with respect to mirroring over
the origin. Therefore, it cannot receive a $g_1^2$ correction, and the
lowest order correction to the estimator must scale as $g_1^3$.  Note that 
in Equation \ref{eq:e1_thetat}, this means that $^{12}\mathbf{W} = 0$.

Therefore, we construct a third-order estimator from quantities $^1\mathbf{U}$
and $^3\mathbf{U}$. Again, because of the symmetry of the problem, we construct
it assuming the problem is one dimensional, i.e. we are attempting to
recover the $g_1$ component. In that case all $\mathbf{W}$ quantities are
scalar.   

Starting with the system of equations:
\begin{eqnarray}
  \left<^1\mathbf{U}\right> &=& ^{11}\mathbf{W} \Delta\bitheta +\frac{^{13}\mathbf{W}}{6}\Delta\bitheta^3 + \ldots, \\
  \left<^3\mathbf{U}\right> &=& ^{31}\mathbf{W} \Delta\bitheta +\frac{^{33}\mathbf{W}}{6}\Delta\bitheta^3 + \ldots, 
\end{eqnarray}
it is not difficult to show that, ignoring higher order terms,
\begin{equation}
  \frac{ ^{33}\mathbf{W} \left<^1\mathbf{U}\right>  - ^{31}\mathbf{W}
    \left<^3\mathbf{U}\right>}{^{11}\mathbf{W}^{33}\mathbf{W} - ^{13}\mathbf{W} ^{31}\mathbf{W}} = \Delta\bitheta
\end{equation}

Hence, we can write an ansatz:
\begin{equation}
 \boldsymbol{E}_3 =   \frac{ ^{33}\mathbf{W} ^1\mathbf{U}  - ^{31}\mathbf{W} ^3\mathbf{U}}{^{11}\mathbf{W}^{33}\mathbf{W} - ^{13}\mathbf{W} ^{31}\mathbf{W}}
\label{eq:e3}
\end{equation}
Since $\mathbf{W}$ quantities do not depend on data, $\left<\boldsymbol{E}_3 \right>
=\Delta\bitheta$ and hence this is our third order estimator.  For more
realistic cases, the rotational symmetry might be broken due to
systematic and instrumental effects and for completeness we show how
to build a complete 3$^{\rm rd}$ order estimator in Appendix
\ref{sec:general-3rm-rd}.

\subsection{Results for toy model}
For this toy example, we can calculate the likelihood and its
derivatives simply by brute force Monte Carlo - we can draw a large
enough number of samples from the parent distribution such that the
gridded values of sampled $\ve$ become a good approximation for the
probability distribution. The derivatives are then calculated by
finite difference methods from gridded likelihoods. Note that this
short-cut is unlikely to work in a more realistic setting due to
the higher dimensionality of the problem.

In Figure \ref{fig:1}, we plot the $i$-th derivative of the likelihood
with respect to $g_1$, that is quantities $P$, $Q_1$, $R_{11}$,
$S_{111}$, showing how the posterior distribution of ellipticities responds
to shear at each order.
 
In Figure \ref{fig:2}, we show results for the three estimators
discussed in this text. As expected, the $\boldsymbol{E}_{BA}$ and $\boldsymbol{E}_1$ estimators
show a quadratic increase in bias as a function of shear, which is
mostly removed by the $\boldsymbol{E}_3$ estimator. In this particular case, our
$\boldsymbol{E}_1$ estimator seems to be performing somewhat better than the
original $\boldsymbol{E}_{BA}$ estimator, although it is not clear whether this
will translate to similar gains in more realistic scenarios. However,
the $\boldsymbol{E}_3$ estimator is designed to be more accurate and performs with
a 0.1\% relative precision all the way to shears of 0.2, at which
point we are well out of the validity of the small shear approximation, and
flexion effects \cite{2005ApJ...619..741G} become important, which are
not captured in this toy model.

In Figure \ref{fig:3}, we show the error (square root of variance) for
the three estimators discussed here, normalized to the Fisher matrix
prediction at zero shear. As we can see, both $\boldsymbol{E}_{BA}$ and $\boldsymbol{E}_1$
converge to the Fisher matrix prediction at zero shear, but $\boldsymbol{E}_3$ is marginally
noisier. The effect is small, sub 1\%, but clearly detectable. For
higher shear, the $\boldsymbol{E}_1$ and $\boldsymbol{E}_3$ estimators begin to become slightly less noisy
than the zero-shear Fisher prediction. Note that this does not violate
the Cramer-Rao bound, since the bound only holds if the true shear is zero.

Finally, we demonstrate explicitly that our estimator can measure
correlations. To that end, we draw pairs of galaxies with shear
$\ibg_a$ and $\ibg_b$, which we randomly choose to follow
\begin{equation}
    \left<\ibg_a \ibg_a^T\right> = \left< \ibg_b \ibg_b^T \right> = \left( \begin{array}{cc}
0.05^2 & 0 \\
 0  & 0.05^2 \\
\end{array} \right)
\end{equation}
and
\begin{equation}
    \left<\ibg_a \ibg_b^T \right> = \left( \begin{array}{cc}
0.00125 & 0.00075 \\
 0.00075  & 0.00125 \\
\end{array} \right).
\end{equation}
These pairs of galaxies are modeled using Equations \ref{eq:toy1},
\ref{eq:toy2}, and \ref{eq:toy3} with $\sigma_p=0.3$, $\sigma_n=0.05$
to obtain observed values and then with the $E_3$ estimator to obtain
an estimate.  These estimates where then used to obtain the
correlations: $\left<\tilde{\ibg}_a \tilde{\ibg}_b^T \right>_{11} =
0.00125319\pm2.8 \times 10^{-6}$ and $\left<\tilde{\ibg}_a
  \tilde{\ibg}_b^T \right>_{12} = 0.007552\pm2.8\times 10^{-6}$,
consistent with the input values and sub-percent level accurate. Of
course, this exercise \emph{had} to work, so it is really just a
sanity check.

\section{Conclusions}

In this paper, we have derived a general framework for generating
unbiased estimators. The framework is general and can be used wherever
we are measuring a quantity which is perturbatively close to the
assumed model. We have shown that the inverse of Fisher matrix
multiplied by the first derivative vector is a general formula for a
first order unbiased estimator.  In special cases such as an optimal
quadratic estimator, the estimator is unbiased at all orders.  We have
applied our framework to the problem of estimating weak lensing shear
and constructed a first and third-order estimator.

In the realm of the toy problem of BA14, our third-order estimator is
unbiased for all relevant shear magnitudes with a negligible increase
in the estimator variance compared to the Fisher prediction at zero
shear. In typical weak-lensing analyses, shears are small enough that
the first-order estimator may be sufficient. However, there are two
cases where third order correction might matter. First, when measuring
the cosmic shear power spectrum, an error term proportional to $g^3$
will ``renormalize'' to give a correction to the measured shear power
spectrum proportional to $\left<\left|\ibg\right|^2\right>P_{gg}$,
where $P_{gg}$ is the true shear power spectrum. This is of the same
order of magnitude as the overall LSST error
\cite{2009arXiv0912.0201L}.  Second, in regions of high-shear, such as
those around clusters of galaxies, the third-order estimator will be
useful, simply because shear are large-enough that the third order
correction matters.  The formalism presented here can trivially be
extended to the flexion measurement, and it should correctly account
for the correlation between shear and flexion. We refrain from making
more quantitative statements since it is not clear how realistic the
toy model is. 

More importantly, we have constructed an estimator which performs as
well as the BA14 estimator, but also returns shear estimates for
individual galaxies, which makes it usable in direct measurements of
the $n$-point function of the shear field. 

We also note that to some extent the main problem with shear
measurements is not the underlying framework, which is the focus of
this paper, but the bias arising from inadequate modeling of the
properties of unlensed galaxies, and it might turn out that these
problems are best solved using very phenomenological approaches as
those discussed in e.g. \cite{2007MNRAS.382..315M,2013arXiv1303.4739R}.

Putting this estimator into practice might be more complicated. In
particular, in its current incarnation, it gives the same weight to
all galaxies, while we know that this will not hold in
reality. The correct way to solve this problem is to separate galaxies
into sub-classes in a way that does not correlate (or negligibly
correlates) with the underlying shear. A separate estimator can be
constructed for each class, and the Fisher matrix is the appropriate
weight. We leave testing of this framework in more realistic settings for
the future work.

\acknowledgments The authors thank Gary Bernstein and Erin Sheldon for
useful conversations.  M.M. is supported by an SBU-BNL Research
Initiatives Seed Grant: Award Number 37298, Project Number
1111593. AS is supported by the DOE Early Career award.

\bibliography{cosmo,cosmo_preprints,dodelson}

\appendix
\section{Example of bias of ML estimator}
\label{sec:biasml}
Here we give a concrete example of a likelihood for which the maximum
likelihood estimator is biased. In general, this happens with
asymmetric likelihoods. Consider:
\begin{equation}
  L=x\lambda^2 e^{-\lambda x},
\end{equation}
where $x>0$ is the ``data'' and $\lambda>0$ is the theory parameter.
Given exactly one measurement $x$, the maximum likelihood estimator
(i.e. the estimator where one would end up upon iterations of
Newton-Raphson steps) is
\begin{equation}
  E_{ML}=\frac{2}{x},
\end{equation}
whose expectation value is $2\lambda$, i.e, wrong by a factor of two.
Expanding around $\lambda=l$, our first order estimator is given by 
\begin{equation}
  E_1=\frac{l(4-lx)}{2}
\end{equation}
which is unbiased up to quadratic order in $\lambda-l$. Interestingly, 
\begin{equation}
  E=\frac{1}{x}
\end{equation}
is unbiased at all orders and is neither ML nor our perturbative estimator.

\section{General 3$^{\rm rd}$ order estimator}
\label{sec:general-3rm-rd}

For completeness we demonstrate how to build a full third order
estimator. This procedure can be trivially generalized to any order.
We write the Equation \eqref{eq:rax} to up to third order in an
``unrolled'' matrix form 
\begin{equation}
\left<U\right> = W g,
\end{equation}

\noindent where we have, assuming that there are two theory parameters that we
want to recover ($g_1$ and $g_2$),
\begin{equation}
U=\left[
\begin{array}{c}
 \left< ^1U_1 \right> \\
 \left< ^1U_2 \right> \\
\left< ^2U_{11} \right> \\
\left< ^2U_{12} \right> \\
\left< ^2U_{22} \right> \\
\left< ^3U_{111} \right> \\
\left< ^3U_{112} \right> \\
\left< ^3U_{122} \right> \\
\left< ^3U_{222} \right> 
 \end{array} 
\right]
\end{equation}
and
\begin{widetext}
\begin{equation}
W=
\left[
\begin{array}{ccccccccc}
 ^{11}W_{1|1} &  ^{11}W_{1|2} & ^{12}W_{1|11} & ^{12}W_{1|12} &  ^{12}W_{1|22} & ^{13}W_{1|111} & ^{13}W_{1|112} & ^{13}W_{1|122} & ^{13}W_{1|222} \\
 ^{11}W_{2|1} &  ^{11}W_{2|2} & ^{12}W_{2|11} & ^{12}W_{2|12} &  ^{12}W_{2|22} & ^{13}W_{2|111} & ^{13}W_{2|112} & ^{13}W_{2|122} & ^{13}W_{2|222} \\
 ^{21}W_{11|1} &  ^{21}W_{11|2} & ^{22}W_{11|11} & ^{22}W_{11|12} &  ^{22}W_{11|22} & ^{23}W_{11|111} & ^{23}W_{11|112} & ^{23}W_{11|122} & ^{23}W_{11|222} \\
 ^{21}W_{12|1} &  ^{21}W_{12|2} & ^{22}W_{12|11} & ^{22}W_{12|12} &  ^{22}W_{12|22} & ^{23}W_{12|111} & ^{23}W_{12|112} & ^{23}W_{12|122} & ^{23}W_{12|222} \\
 ^{21}W_{22|1} &  ^{21}W_{22|2} & ^{22}W_{22|11} & ^{22}W_{22|12} &  ^{22}W_{22|22} & ^{23}W_{22|111} & ^{23}W_{22|112} & ^{23}W_{22|122} & ^{23}W_{22|222} \\
 ^{31}W_{111|1} &  ^{31}W_{111|2} & ^{32}W_{111|11} & ^{32}W_{111|12} &  ^{32}W_{111|22} & ^{33}W_{111|111} & ^{33}W_{111|112} & ^{33}W_{111|122} & ^{33}W_{111|222} \\
 ^{31}W_{112|1} &  ^{31}W_{112|2} & ^{32}W_{112|11} & ^{32}W_{112|12} &  ^{32}W_{112|22} & ^{33}W_{112|111} & ^{33}W_{112|112} & ^{33}W_{112|122} & ^{33}W_{112|222} \\
 ^{31}W_{122|1} &  ^{31}W_{122|2} & ^{32}W_{122|11} & ^{32}W_{122|12} &  ^{32}W_{122|22} & ^{33}W_{122|111} & ^{33}W_{122|112} & ^{33}W_{122|122} & ^{33}W_{122|222} \\
 ^{31}W_{222|1} &  ^{31}W_{222|2} & ^{32}W_{222|11} & ^{32}W_{222|12} &  ^{32}W_{222|22} & ^{33}W_{222|111} & ^{33}W_{222|112} & ^{33}W_{222|122} & ^{33}W_{222|222} \\
  \end{array} 
 \right]
\end{equation}
\end{widetext} 
and 
\begin{equation}
g=
\left[
\begin{array}{c}
g_1\\
g_2\\
g_1g_1\\
2\times g_1g_2\\
g_2g_2\\
g_1g_1g_1\\
3\times g_1g_1g_2\\
3\times g_1g_2g_2\\
g_2g_2g_2\\
\end{array}
\right].
\end{equation}
In expression for $W$, we have used a pipe symbol to separate indices corresponding to
the left and right sides of the equation. Solving this matrix equation for the vector $g$. 
We have
\begin{equation}
  g=W^{-1} \left<U\right>
\end{equation}
We can now write an ansatz for the estimator:
\begin{equation}
  E = W^{-1} U
\end{equation}
Since $W$ does not depend on data, it trivially follows that
\begin{equation}
\left<E \right> = W^{-1}\left<U\right> = g 
\end{equation}
Hence, the first two components of $E$, namely $E_1$ and $E_2$ are
unbiased estimators for the first two components of $g$, that is $g_1$
and $g_2$. In other words, the linear algebra has given us the
particular \emph{linear} combination of $U$ quantities which average
to $g_1$ and $g_2$ without any contribution from terms quadratic and
cubic in $g$.

\end{document}